\documentclass[twocolumn,showpacs,preprintnumbers,amsmath,amssymb]{revtex4}

\usepackage{graphicx}
\usepackage{dcolumn}
\usepackage{bm}

\newcommand{\ps}{p\!\!\!/}

\begin{document}

\preprint{}

\title{Exact semi-relativistic model for ionization of atomic hydrogen by electron impact}
\author{Y. Attaourti}
\email{attaourti@ucam.ac.ma} \affiliation{ Laboratoire de
Physique des Hautes Energies et d'Astrophysique, Facult\'e des
Sciences Semlalia, Universit\'e Cadi Ayyad Marrakech, BP : 2390,
Morocco.}
\author{S. Taj }\email{souad_taj@yahoo.fr}\author{B. Manaut}
\affiliation{UFR de Physique Atomique Mol\'eculaire et Optique
Appliqu\'ee,
 Facult\'e des Sciences, Universit\'e Moulay
Isma\"{\i}l, BP : 4010, Beni M'hamed, Mekn\`es, Morocco.}

\begin{abstract}
We present a semi-relativistic model for the description of the
ionization process of atomic hydrogen by electron impact in the
first Born approximation by using the Darwin wave function to
describe the bound state of atomic hydrogen and the
Sommerfeld-Maue wave function to describe the ejected electron.
This model, accurate to first order in $Z/c$ in the relativistic
correction, shows that, even at low kinetic energies of the
incident electron, spin effects are small but not negligible.
These effects become noticeable with increasing incident electron
energies. All analytical calculations are exact and our
semi-relativistic results are compared with the results obtained
in the non relativistic Coulomb Born Approximation both for the
coplanar asymmetric and the binary coplanar geometries.
\end{abstract}

\maketitle
\section{Introduction}
Relativistic $(e,2e)$ processes have been reviewed both from the
experimental and theoretical point of view [1]. As one deals with
atomic hydrogen, the value of the parameter $Z\alpha$ is very
lower than one, where $Z$ is the atomic charge number and $\alpha$
is the fine structure constant. Therefore, it is convenient and
sufficient to use approximate wave functions of a mathematically
simpler structure than the exact analytical wave functions needed
to describe relativistic $(e,2e)$ processes. A numerical approach
to an exact description of the relativistic ionization of atomic
hydrogen by electron impact could be carried out but we will focus
instead on an alternative approach that will give nearly the same
results as the exact description if the condition $Z\alpha\ll 1$
is satisfied. In $(e,2e)$ processes, relativistic effects are
important and all electrons (the incident, scattered and ejected)
can have very high velocities. One has to consider many
interactions (to name some, retardation interaction, magnetic
interaction and spin-dependent interaction). For atomic hydrogen,
many experimental and theoretical contributions have been made
[2-3]. Some were successful but the theoretical situation for all
set-ups and kinematics is far from resolved, at least
analytically. Many calculations have resorted to various
approximations. For example, plane wave models [4-7] are
successful in the coplanar binary geometries [4] and for fast
scattered and ejected electrons. The first Born approximation
(FBA) has been used to describe asymmetric geometries at non
relativistic energies [8-9]. In this approximation, the incident
and scattered electrons are described by plane waves whereas the
ejected electron is treated as a Coulomb wave. Many authors
extended this approximation to the relativistic domain. Das et al
[10-11] employed a semi-relativistic Sommerfeld-Maue wave
functions to describe the ejected electron. Jakuba\ss a-Amundsen
evaluated the first-order transition matrix element $S_{fi}$ using
semi-relativistic Coulomb wave functions times a free spinor i.e
neglecting the relativistic contraction of the bound state and
approximating the continuum Coulomb state by a relativistic
Coulomb wave times a free spinor. This model did well in
predicting integrated cross-sections [12] but yielded a value for
the absolute triple differential cross section (TDCS) too large.
For the Coulomb approximation, Jakuba\ss a-Amundsen argued that
one could not neglect the Coulomb potential in the treatment of
inner-shell ionization of high-$Z$ atoms. Agreement with
experiment was encouraging for intermediate values of $Z$. The
merits and shortcomings of this theory have been analyzed in [14].
Thereafter, a fully relativistic version was produced [15] which
showed that the original physical insight was essentially
correct.\\

 In this contribution, we present a theoretical semi-relativistic
model, the semi-relativistic Coulomb Born Approximation (SRCBA) in
a closed and exact form for the description of the ionization of
atomic hydrogen by electron impact that is valid for all
geometries. Indeed, in the non relativistic Coulomb Born
Approximation (NRCBA), a well known integral occurs [16] and is
usually denoted by $I(\lambda )$. In this article, we show that
the main contribution to the unpolarized triple differential cross
section (TDCS) in the SRCBA corresponding to the ionization of
atomic hydrogen in its ground state by electron impact comes from
this term added to relativistic corrections valid to first order
in $Z/c$. These relativistic corrections contain a new integral we
have denoted $J(\lambda )$ and in the Appendix, we give the formal
derivation of this integral. To our knowledge, it is the first
time that such an integral is written down analytically. Needless
to say that all numerical appropriate tests to check the validity
of the analytical result we have found have been carried out with
a very good degree of accuracy. It turns out that spin effects can
be accounted for even at low kinetic energies of the incident
electron in the case of the Ehrhardt coplanar asymmetric geometry
[17] where, for a given kinetic energy $T_{i}$ of the fast
incident electron, a fast (''scattered '') electron of kinetic
energy $T_{f}$  is detected in coincidence with a slow
(''ejected'') electron of kinetic energy $T_{B}$. These spin
effects as well as the relativistic effects become noticeable with
increasing incident electron
kinetic energy.\\

The organization of this paper is as follows : in section II, we
present the semi relativistic formalism of $(e,2e)$ reaction and
give a detailed account of the various terms that contribute to
the unpolarized TDCS, in section III, we discuss the results we
have obtained and we end by a brief conclusion in section IV. The
formal derivation of the integral $J(\lambda )$ is given in the
Appendix. Throughout this work, atomic units ($a.u$) are used
$(\hbar =m_{e}=e=1)$ where $m_{e}$ is the electron rest mass.

\section{The unpolarized triple differential cross section}

In this section, we calculate the exact analytical expression of
the semi relativistic unpolarized TDCS in the SRCBA corresponding
to the ionization of atomic hydrogen by electron impact. The
transition matrix element for the direct channel (exchange
effects are neglected) is given by
\begin{eqnarray}
S_{fi} &=&-i\int dt<\psi _{p_{f}}(x_{1})\phi _{f}(x_{2})\mid
V_{d}\mid \psi _{p_{i}}(x_{1})\phi _{i}(x_{2})>
 \nonumber \\
 &=&-i\int_{-\infty }^{+\infty }dt\int d\mathbf{r}_{1}\overline{\psi }%
_{p_{f}}(t,\mathbf{r}_{1})\gamma _{(1)}^{0}\psi
_{p_{i}}(t,\mathbf{r}_{1}) \nonumber \\ && \quad \quad\times <\phi
_{f}(x_{2})\mid V_{d}\mid \phi _{i}(x_{2})> \label{1}
\end{eqnarray}
In Eq. (1), $V_{d}$ is the direct interaction potential :

\begin{equation}
V_{d}=\frac{1}{r_{12}}-\frac{1}{r_{1}}  \label{2}
\end{equation}
$\mathbf{r}_{1}$ are the coordinates of the incident and
scattered electron, $\mathbf{r}_{2}$ are the atomic electron
coordinates, $r_{12}=$ $\mid \mathbf{r}_{1}-\mathbf{r}_{2}\mid $
and $r_{1}=\mid \mathbf{r}_{1}\mid $.
The wave function $\psi _{p_{i}}(x_{1})=\psi _{p}(t,\mathbf{r}%
_{1})=u(p,s)\exp (-ip.x)/\sqrt{2EV\text{ }}$ is the electron wave
function described by a free Dirac spinor normalized to the
volume $V$ and $\phi _{i,f}(x_{2})=\phi _{i,f}(t,\mathbf{r}_{2})$
are the semi relativistic wave functions of the hydrogen atom
where the index $i$ stands for the initial state, namely the
ground state and the index $f$ stands for the final state. The
quantity $p.x=p_{\mu }x^{\mu }$ is the Lorentz scalar product.
The semi relativistic wave function of the hydrogen atom used is
the Darwin wave function for bound states [18]

\begin{equation}
\phi _{i}(t,\mathbf{r}_{2})=\exp (-i\varepsilon _{b}t)\varphi ^{(\pm )}(%
\mathbf{r}_{2})  \label{3}
\end{equation}
where

\begin{equation}
\varphi ^{(\pm
)}(\mathbf{r}_{2})=(\mathsf{1}_{4}-\frac{i}{2c}\mathbf{\alpha
.\nabla }_{(2)})u^{(\pm )}\varphi _{0}(\mathbf{r}_{2})  \label{4}
\end{equation}
is a quasi relativistic bound state wave function accurate to
first order in $Z/c$ in the relativistic corrections (and
normalized to the same order) with $\varphi _{0}$ being the non
relativistic bound state hydrogenic function. The spinors
$u^{(\pm )}$ are such that $u^{(+)}=(1,0,0,0)^{T}$ and
$u^{(-)}=(0,1,0,0)^{T}$ and represent the basic four-component
spinors for a particle at rest with spin-up and spin-down,
respectively. For the spin up, we have

\begin{eqnarray}
\varphi ^{(+)}(\mathbf{r}_{2})&=&N_{D} \left(
\begin{array}{c}
 1 \\
  0 \\
  \frac{i}{2c}\cos \theta _{2} \\
  \frac{i}{2c}\sin (\theta _{2})\exp (i\phi _{2})
\end{array}\right)
\frac{1} {\sqrt{\pi }} e^{-r_{2}}
\end{eqnarray}
and for the spin down, we have

\begin{eqnarray}
\varphi ^{(-)}(\mathbf{r}_{2})&=& N_{D}\left(
\begin{array}{c}
 0 \\
  1 \\
  \frac{i}{2c}\sin
(\theta _{2})\exp (-i\phi _{2}) \\
  -\frac{i}{2c}\cos
(\theta _{2})
\end{array}\right)
\frac{1} {\sqrt{\pi }} e^{-r_{2}}
\end{eqnarray}
where

\begin{equation}
N_{D}=2c/\sqrt{4c^{2}+1}  \label{7}
\end{equation}
is a normalization constant lower but very close to one. The wave function $%
\phi _{f}(t,\mathbf{r}_{2})$ in Eq. (1) is the Sommerfeld-Maue
wave function for continuum states [18] also accurate to the order
$Z/c$ in the relativistic corrections. We have $\phi
_{f}(t,\mathbf{r}_{2})=\exp (-iE_{b}t)\psi
_{p_{f}}^{(-)}(\mathbf{r}_{2})$ and

\begin{eqnarray}
&&\psi _{p_{f}}^{(-)}(\mathbf{r}_{2}) =\exp (\pi \eta
_{B}/2)\Gamma (1+i\eta _{B}) \exp
(i\mathbf{p}_{B}.\mathbf{r}_{2})\nonumber \\
 &\times&\big\{\mathsf{1}_{4}-\frac{ic}{2E_{B}}%
\mathbf{\alpha} .\mathbf{\nabla }_{(2)}\big\}\, _{1}F_{1}(-i\eta
_{B},1,-i(p_{B}r_{2}+\mathbf{p}_{B}.\mathbf{r}_{2}))\nonumber \\
&\times&\frac{%
u(p_{B},s_{B})}{\sqrt{2E_{B}V}}  \label{8}
\end{eqnarray}
normalized to the volume $V$. The Sommerfeld parameter is given by

\begin{equation}
\eta _{B}=\frac{E_{B}}{c^{2}p_{B}}  \label{9}
\end{equation}
In Eqs. (3) and (8), $\mathbf{\alpha}$ is related to the
$\mathbf{\gamma }$ Dirac matrices [19] and in the standard
representation reads

\begin{equation}
\mathbf{\alpha}=\left(
\begin{array}{ll}
0 & \mathbf{\sigma } \\
\mathbf{\sigma } & 0
\end{array}
\right)  \label{10}
\end{equation}
with $\mathbf{\sigma =}(\sigma _{x},\sigma _{y},\sigma _{z})$ and
the matrices $\sigma _{x},\sigma _{y},\ \sigma _{z}$ are the
usual Pauli matrices. The matrix differential operator
$\mathbf{\mathbf{\alpha} .\mathbf{\nabla} }$ is given by

\begin{equation}
\mathbf{\mathbf{\alpha} .\mathbf{\nabla} =}\left(
\begin{array}{llll}
  0& 0 & \partial _{z} & \partial _{x}-i\partial _{y} \\
  0 & 0 & \partial _{x}+i\partial _{y} & -\partial _{z} \\
  \partial _{z} & \partial _{x}-i\partial _{y}& 0 & 0 \\
 \partial _{x}+i\partial _{y} & -\partial _{z} & 0 & 0
\end{array}
\right)  \label{11}
\end{equation}

We give the final compact form of the Sommerfed-Maue wave function
\begin{widetext}
\begin{eqnarray}
\psi _{p_{f}}^{(-)}(\mathbf{r}_{2}) =\exp (\pi \eta _{B}/2)\Gamma
(1+i\eta _{B})\exp (i\mathbf{p}_{B}.\mathbf{r}_{2})  \Big\{
 \,_{1}F_{1}(-i\eta
_{B},1,-i(p_{B}r_{2}+\mathbf{p}_{B}.\mathbf{r}_{2})
&&+\frac{i}{2cp_{B}}(\mathbf{\alpha .p}_{B}+p_{B}\mathbf{\alpha .}\widehat{%
\mathbf{r}}_{2})\nonumber\\ \times _{1}F_{1}(-i\eta _{B}+1,2,-i(p_{B}r_{2}+\mathbf{p}_{B}.%
\mathbf{r}_{2}))\Big\} \frac{u(p_{B},s_{B})}{\sqrt{2E_{B}V}}
\label{12}
\end{eqnarray}
In Eq. (12), the operator $\mathbf{\alpha .p}_{B}$ acts on the free spinor $%
u(p_{B},s_{B})$ and the operator $\mathbf{\alpha
.}\widehat{\mathbf{r}}_{2}$ acts on the spinor part of the Darwin
function. The direct transition matrix element in Eq. (1) becomes
\begin{eqnarray}
S_{fi} &=&-i\int d\mathbf{r}\frac{\overline{u}(p_{f},s_{f})}{\sqrt{2E_{f}V}}%
\gamma
_{(1)}^{0}\frac{\overline{u}(p_{B},s_{B})}{\sqrt{2E_{B}V}}\gamma
_{(2)}^{0}\Big \{\,_{1}F_{1}(i\eta _{B},1,i(p_{B}r+\mathbf{p}_{B}.\mathbf{r})%
\mathsf{1}_{4}
-\frac{i}{2cp_{B}}(\mathbf{\alpha .p}_{B}+p_{B}\mathbf{\alpha .}\widehat{%
\mathbf{r}})\nonumber\\ &\times&_{1}F_{1}(i\eta
_{B}+1,2,i(p_{B}r+\mathbf{p}_{B}.\mathbf{r}))\Big\} \varphi
^{(\pm )}(\mathbf{r})
\exp (-i\mathbf{p}_{B}.\mathbf{r})[\exp (i%
\mathbf{\Delta .r})-1]\nonumber\\ &\times& \frac{8\pi^{2} }{\Delta
^{2}} \delta (E_{f}+E_{B}-E_{i}-\varepsilon _{b})
\frac{u(p_{i},s_{i})}{\sqrt{2E_{i}V}}\exp (\pi \eta _{B}/2)\Gamma
(1-i\eta _{B})  \label{13}
\end{eqnarray}
This transition matrix element contains three terms, one of which
is given by
\begin{eqnarray}
S_{fi}^{(1)} &=&-i\int d\mathbf{r}\frac{\overline{u}(p_{f},s_{f})}{\sqrt{%
2E_{f}V}}\gamma
_{(1)}^{0}\frac{\overline{u}(p_{B},s_{B})}{\sqrt{2E_{B}V}}\gamma
_{(2)}^{0}\Big\{\,_{1}F_{1}(i\eta
_{B},1,i(p_{B}r+\mathbf{p}_{B}.\mathbf{r}))\Big\}
\varphi ^{(\pm )}(\mathbf{r})\nonumber\\&\times&\exp (-i\mathbf{p}_{B}.\mathbf{r})[\exp (i%
\mathbf{\Delta .r})-1]\frac{8\pi^2 }{\Delta ^{2}}\delta
(E_{f}+E_{B}-E_{i}-\varepsilon _{b})
\frac{u(p_{i},s_{i})}{\sqrt{2E_{i}V}}\exp (\pi \eta _{B}/2)\Gamma
(1-i\eta _{B})  \label{14}
\end{eqnarray}
This term can be recast in the form :

\begin{eqnarray}
S_{fi}^{(1)} &=&-i\left[ H_{1}(\mathbf{q}=\mathbf{\Delta }-\mathbf{p}%
_{B})-H_{1}(\mathbf{q}=-\mathbf{p}_{B})\right]   \frac{\overline{u}(p_{f},s_{f})}{\sqrt{2E_{f}V}}\gamma _{(1)}^{0}\frac{%
\overline{u}(p_{B},s_{B})}{\sqrt{2E_{B}V}}\gamma _{(2)}^{0}\frac{u(p_{i},s_{i})}{\sqrt{%
2E_{i}V}} \frac{8\pi^2 }{\Delta ^{2}}\nonumber\\
&\times&\delta (E_{f}+E_{B}-E_{i}-\varepsilon _{b})\exp (\pi \eta
_{B}/2)\Gamma (1-i\eta _{B}) \label{15}
\end{eqnarray}
In the above expression, $H_{1}(\mathbf{q})$ is given by

\begin{eqnarray}
H_{1}(\mathbf{q})=\int d\mathbf{r}\exp (i\mathbf{q.r})
\,_{1}F_{1}(i\eta _{B},1,i(p_{B}r+\mathbf{p}_{B}.\mathbf{r}))
\varphi ^{(\pm )}(\mathbf{r}) \label{16}
\end{eqnarray}
For instance, if one considers $\varphi ^{(+)}(\mathbf{r})$, the quantity $%
H_{1}(\mathbf{q})$ is given by

\begin{equation}
H_{1}(\mathbf{q})=(I_{1},I_{2},I_{3},I_{4})^{T}  \label{17}
\end{equation}
and one has to evaluate

\begin{eqnarray}
I_{1}=\frac{1}{\sqrt{\pi }}\int d\mathbf{r}\exp (i\mathbf{q.r})\frac{e^{-r}}{%
r}\text{ }_{1}F_{1}(i\eta
_{B},1,i(p_{B}r+\mathbf{p}_{B}.\mathbf{r})) \label{18}
\end{eqnarray}
To do that, we introduce the well-known integral [16]
\begin{eqnarray}
I(\lambda ) =\int d\mathbf{r}\exp (i\mathbf{q.r})\frac{e^{-\lambda r}}{r}%
\text{ }_{1}F_{1}(i\eta
_{B},1,i(p_{B}r+\mathbf{p}_{B}.\mathbf{r}))
=\frac{4\pi }{q^{2}+\lambda ^{2}}\exp \left[ i\eta _{B}\ln (\frac{%
q^{2}+\lambda ^{2}}{q^{2}+\lambda ^{2}+2\mathbf{q.p}_{B}-2i\lambda p_{B}}%
)\right]  \label{19}
\end{eqnarray}
The other integrals can be obtained by noting that
\begin{equation}
\cos \theta \exp (i\mathbf{q.r})=-\frac{i}{r}\frac{\partial }{\partial q_{z}}%
\exp (i\mathbf{q.r})  \label{20}
\end{equation}
and
\begin{equation}
\sin \theta \exp (i\phi )\exp (i\mathbf{q.r})=-\frac{i}{r}(\frac{\partial }{%
\partial q_{x}}+i\frac{\partial }{\partial q_{y}})\exp (i\mathbf{q.r})
\label{21}
\end{equation}
The second term in the transition amplitude given in Eq. (13) is
\begin{equation}
S_{fi}^{(2)}=S_{fi}^{(2),1}+S_{fi}^{(2),2}  \label{22}
\end{equation}
with
\begin{eqnarray}
S_{fi}^{(2),1} &=&-\int d\mathbf{r}\frac{\overline{u}(p_{f},s_{f})}{\sqrt{%
2E_{f}V}}\gamma _{(1)}^{0}\frac{u(p_{i},s_{i})}{\sqrt{2E_{i}V}}\frac{1}{%
2cp_{B}}\frac{\overline{u}(p_{B},s_{B})}{\sqrt{2E_{B}V}}\gamma
_{(2)}^{0}[\gamma _{(2)}^{0}\frac{E_{B}}{c}-\ps_{B}]\varphi ^{(\pm )}(\mathbf{r%
})
_{1}F_{1}(i\eta _{B}+1,2,i(p_{B}r+\mathbf{p}_{B}.\mathbf{r})) \nonumber \\&\times&\exp (-i%
\mathbf{p}_{B}.\mathbf{r})[\exp (i\mathbf{\Delta .r})-1] \exp
(\pi \eta _{B}/2)\Gamma (1-i\eta _{B})\frac{8\pi^2 }{\Delta
^{2}}\delta (E_{f}+E_{B}-E_{i}-\varepsilon _{b})  \label{23}
\end{eqnarray}
and
\begin{eqnarray}
S_{fi}^{(2),2} &=&-\int d\mathbf{r}\frac{\overline{u}(p_{f},s_{f})}{\sqrt{%
2E_{f}V}}\gamma _{(1)}^{0}\frac{u(p_{i},s_{i})}{\sqrt{2E_{i}V}}\frac{1}{2c}%
\frac{\overline{u}(p_{B},s_{B})}{\sqrt{2E_{B}V}}\gamma
_{(2)}^{0}\varphi ^{\prime (\pm )}(\mathbf{r})
_{1}F_{1}(i\eta _{B}+1,2,i(p_{B}r+\mathbf{p}_{B}.\mathbf{r}))\nonumber \\&\times&\exp (-i%
\mathbf{p}_{B}.\mathbf{r})[\exp (i\mathbf{\Delta .r})-1] \exp
(\pi \eta _{B}/2)\Gamma (1-i\eta _{B})\frac{8\pi^2 }{\Delta
^{2}}\delta (E_{f}+E_{B}-E_{i}-\varepsilon _{b})  \label{24}
\end{eqnarray}
In Eq. (24), $\varphi ^{\prime (+)}(\mathbf{r})$ for spin-up is
given by

\begin{eqnarray}
\varphi ^{\prime (+)}(\mathbf{r})&=&N_{D}\left(
\begin{array}{c}
 i/2c \\
  0 \\
  \cos (\theta ) \\
  \sin(\theta )e^{i\phi }
\end{array}\right )
\frac{1}{\sqrt{\pi }}%
e^{-r}\label{25}
\end{eqnarray}
Using the standard procedures of QED [19], one obtains for the
unpolarized TDCS

\begin{equation}
\frac{d\overline{\sigma }}{dE_{B}d\Omega _{B}d\Omega _{f}}=\frac{1}{2}%
\sum_{s_{i},s_{f}}\sum_{s_{B}}\frac{1}{2}\sum_{s_{t}}\frac{d\sigma }{%
dE_{B}d\Omega _{B}d\Omega _{f}}  \label{26}
\end{equation}
evaluated for $E_{f}=E_{i}+\varepsilon _{b}-E_{B}$, where $%
\sum_{s_{t}}(...)/2$ denotes the averaged sum over the spin
states of the target atomic hydrogen with
\begin{eqnarray}
\frac{d\sigma }{dE_{B}d\Omega _{B}d\Omega _{f}}=\frac{1}{64c^{6}\pi ^{3}}%
\frac{|p_{f}||p_{B}|}{|p_{i}|}\frac{\exp (\pi \eta _{B})}{\Delta
^{4}}\left| \Gamma (1-i\eta _{B})\right| ^{2}\left| \widetilde{S}_{fi}^{(1)}+\widetilde{S%
}_{fi}^{(2),1}+\widetilde{S}_{fi}^{(2),2}\right| ^{2}  \label{27}
\end{eqnarray}
To our knowledge, in the expressions of $\widetilde{S}_{fi}^{(2),1}$ and $%
\widetilde{S}_{fi}^{(2),2}$, a new integral occurs. We have
calculated this integral analytically. Details of its derivation
are given in the Appendix. This integral is
\begin{eqnarray}
J(\lambda ) =\int d\mathbf{r}\exp (i\mathbf{q.r})\frac{e^{-\lambda r}}{r}%
\text{ }_{1}F_{1}(i\eta
_{B}+1,2,i(p_{B}r+\mathbf{p}_{B}.\mathbf{r})) =\frac{4\pi }{(q^{2}+\lambda ^{2})} \text{} _{2}F_{1}(i\eta _{B}+1,1,2,-2%
\frac{(\mathbf{q.p}_{B}-i\lambda p_{B})}{q^{2}+\lambda ^{2}})
\label{28}
\end{eqnarray}
All the calculations in Eq. (27) can be done analytically and only
five terms from nine are non zero, the diagonal terms $\left| \widetilde{S}%
_{fi}^{(1)}\right| ^{2}$, $\left| \widetilde{S}_{fi}^{(2),1}\right| ^{2}$, $%
\left| \widetilde{S}_{fi}^{(2),2}\right| ^{2}$ and $\widetilde{S}%
_{fi}^{(1)\dagger }\widetilde{S}_{fi}^{(2),1}$ as well as $\widetilde{S}%
_{fi}^{(2),1\dagger }\widetilde{S}_{fi}^{(1)}$. In Eq. (26), the
different sums over spin states give rise to the following results
\begin{eqnarray}
\left.
\begin{array}{c}
\frac{1}{2}\sum_{s_{i},s_{f}}\left|
\overline{u}(p_{f},s_{f})\gamma
_{(1)}^{0}u(p_{i},s_{i})\right| ^{2}=2c^{2}
(\frac{2E_{i}E_{f}}{c^{2}}-(p_{i}.p_{f})+c^{2}) \\
\sum_{s_{B}}\left| (\overline{u}(p_{B},s_{B})\gamma
_{(2)}^{0}[\gamma
_{(2)}^{0}\frac{E_{B}}{c}-\ps_{B}]\right| ^{2}=4E_{B}(\frac{E_{B}^{2}}{c^{2}}%
-c^{2}) \\ \sum_{s_{B}}\left| (\overline{u}(p_{B},s_{B})\gamma
_{(2)}^{0}\right| ^{2}=4E_{B} \\ \frac{1}{2}\sum_{s_{t}}(....)=1
\end{array}
\right.  \label{29}
\end{eqnarray}
\end{widetext}
\section{Results and discussion}
\subsection{Coplanar asymmetric geometries}
We begin our discussion by
considering well known results in the non relativistic domain,
namely the results of Byron and Joachain [17] and those of
Berakdar [21].
\begin{figure}[h] \centering
\includegraphics[angle=0,width=8.5cm,height=5.8cm]{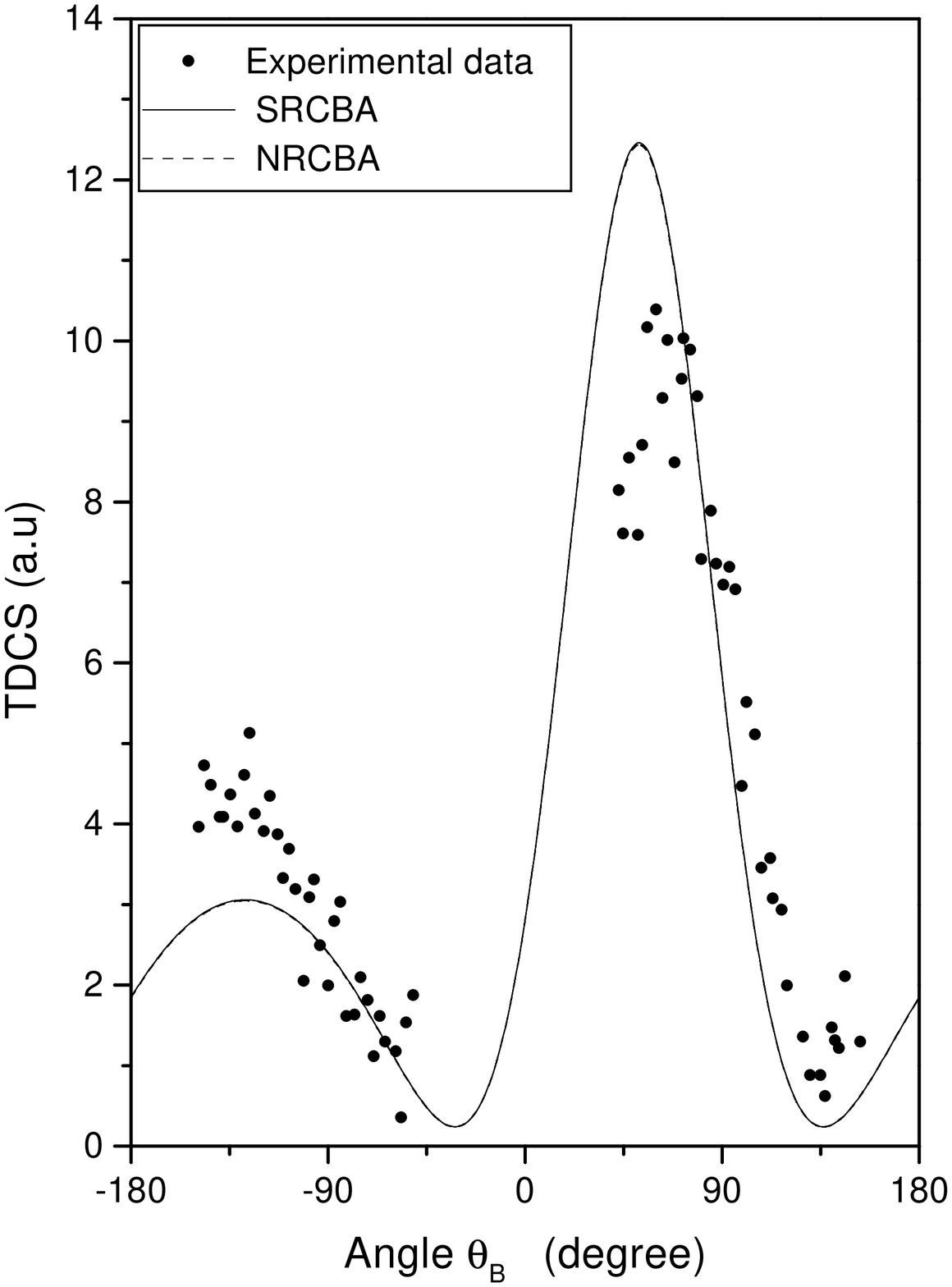}
\caption{The two TDCSs. The solid line represents the
relativistic TDCS in the semi-relativistic Coulomb Born
approximation, the long-dashed line represents the corresponding
TDCS in the non relativistic Coulomb Born approximation. The
incident electron kinetic energy is $T_i=250\quad eV$ and the
ejected electron kinetic energy is $T_B=5\quad eV$. Experimental
data is from \cite{24}}.
\end{figure}All these results are obtained in
the coplanar asymmetric geometry. Let us consider the process
whereby an incident electron with a kinetic energy $E_i=250\text{
}eV$ scatters with a hydrogen atom. The ejected electron is
observed to have a kinetic energy $E_B=5\text{ } eV$ and the
scattered electron is observed having an angle $\theta_f=3^{°}$.
In this particular case, the CBA is not as accurate as the results
obtained within the framework of the Eikonal Born series [17]
which contains higher order corrections.

 Nevertheless, as it can
be seen in Fig. 1, the agreement between the non relativistic and
semi-relativistic results is good since we obtain two identical
curves.
\begin{figure}[h] \centering
\includegraphics[angle=0,width=7cm,height=5.8cm]{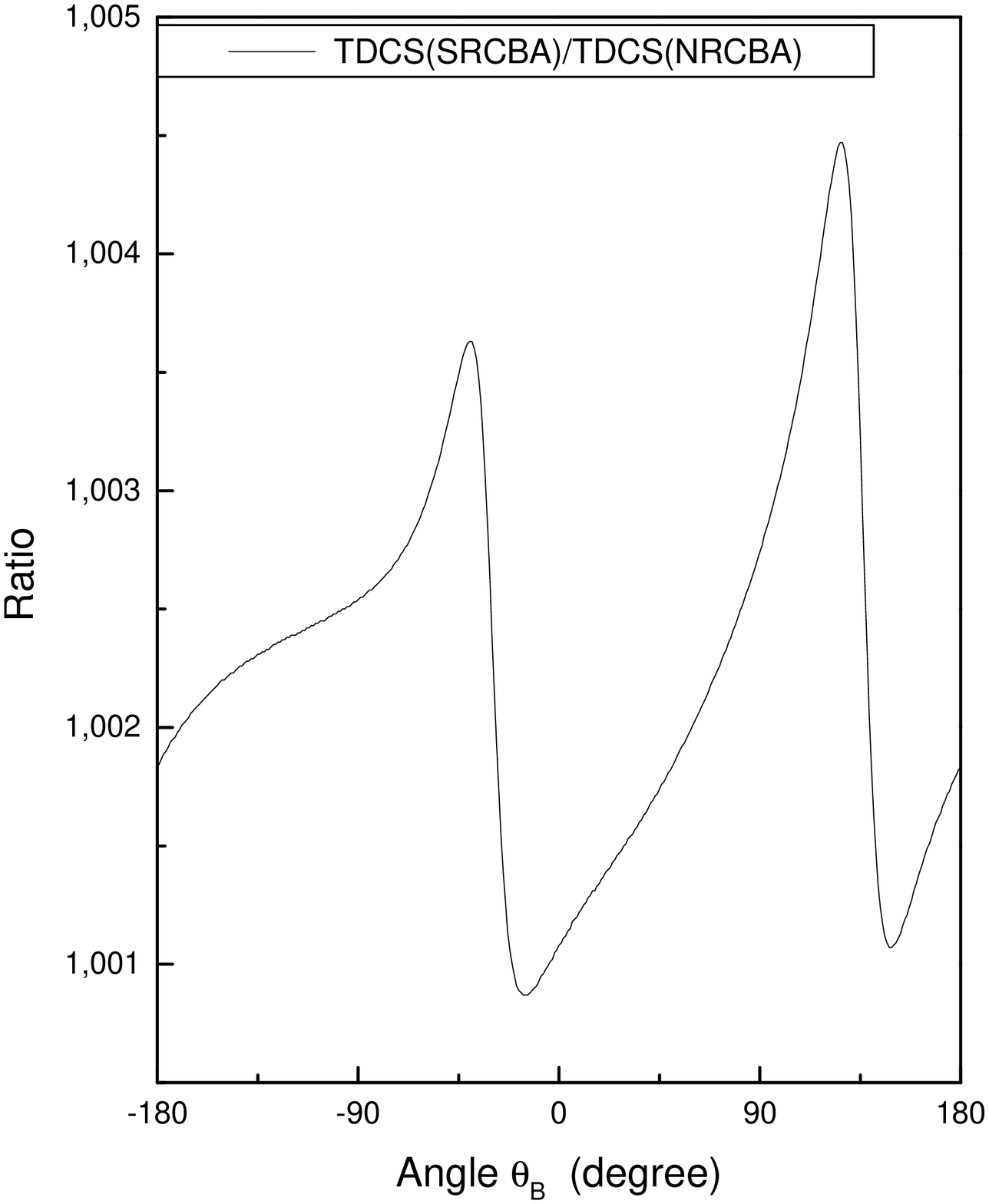}
\caption{The ratio  $TDCS(SRCBA)/TDCS(NRCBA)$ as function of the
angle $\theta_{B}$ with $\theta_{f}=3^{\circ}$. The incident
electron kinetic energy is $T_i=250\quad eV$ and the ejected
electron kinetic energy is $T_B=5\quad eV$}.
\end{figure} However, even in this non relativistic regime,
small effects due to the semi-relativistic treatment of the wave
functions we have used, show that there are indeed small effects
that can only be tracked back to the spin. Indeed, if we plot the
ratio of the semi-relativistic TDCS and the non relativistic TDCS,
it emerges that however small, these spin effects can reach $0,45
\%$ for some specific angles. We recall that the TDCS has extrema,
in particular when the direction of $\mathbf{p_B}$
 coincides with that of the vectors $\mathbf{\Delta}$ and
$-\mathbf{\Delta}$ and this can be seen in Fig. 2.

 In the former
case, the extremum is always a maximum and in the latter case the
extremum is a local maximum. The two TDCSs exhibit in this
geometry a forward or binary peak with a maximum in the direction
of $\mathbf{\Delta}$ and a recoil or backward peak in the opposite
direction $-\mathbf{\Delta}$. The locations of such extrema are
$\theta_B\approx -128^{\circ}$ with a ratio equal to $1.00234$ and
$\theta_B\approx 52^{\circ}$ with a ratio equal to $1.00185$.
These mechanisms for the emergence of the binary-recoil peak
structure are also present even when one uses the simplest
description in which plane waves for incoming and outgoing
particles are assumed [22].

Now, if we compare our result with the result obtained by Berakdar
[21], we also obtain a good agreement. But before beginning the
discussion proper, let us recall the formalism used by Berakdar.
His calculations were performed within a model where the
three-body final state is described by a product of three
symmetrical, Coulomb-type functions. Each of these functions
describes the motion of a particular two-body subsystem in the
presence of a third charged particle. Thereafter, he made a
comprehensive comparison with available experimental data  and
with other theoretical models. He ended his study by concluding
that generally, good agreement is found with the absolute
measurements but that however, in some cases discrepancies between
various theoretical predictions and experimental findings are
obvious, which highlights the need for a theoretical and
experimental benchmark study of these reactions.
\begin{figure}[ht]
\centering
\includegraphics[angle=0,width=7cm,height=5.8cm]{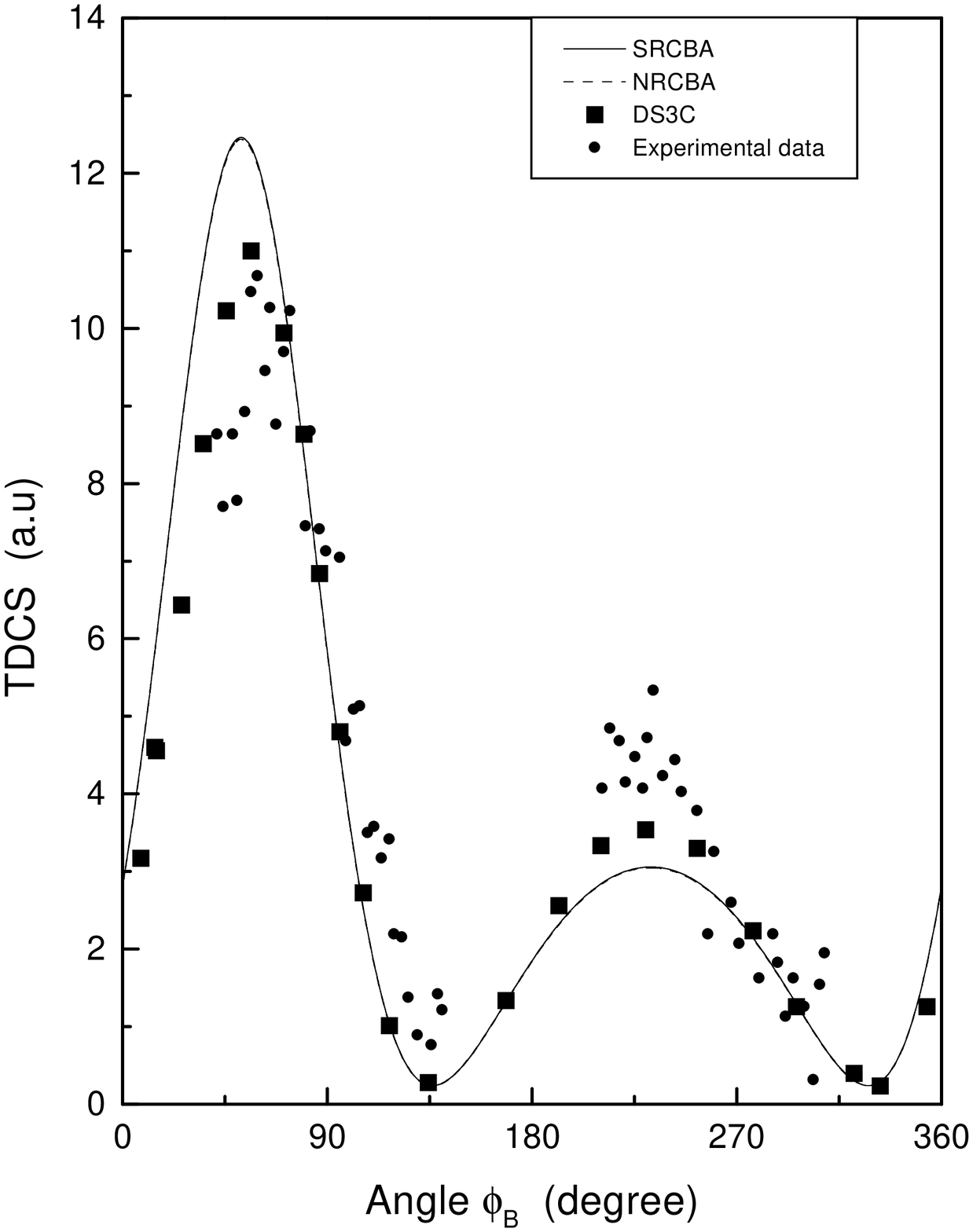}
\caption{The two TDCSs. The solid line represents the
relativistic TDCS in the semi-relativistic Coulomb Born
approximation, the long-dashed line represents the corresponding
TDCS in the non relativistic Coulomb Born approximation, the
symbols square and circle respectively represent the formalism of
DS3C and the experimental data. We keep the same energies as in
Fig. 1.  Experimental data is from \cite{9}}.
\end{figure}
In Fig. 3, we compare our results with those obtained by Berakdar
for an incident electron kinetic energy $E_{i}= 250\text{ }eV$
for the case of a coplanar asymmetric geometric where
$\theta_{f}=\theta_{B}=90^{\circ}$. The ejected electron kinetic
energy is $E_{B}= 5\text{ }eV$ and $\phi_{f}=357^{\circ}$. What is
remarkable is the agreement between our results and his bearing in
mind that he used the DS3C formalism (DS3C stands for dynamical
screening theory with three Coulomb-type functions). Another
atypical result related to our calculations is the behavior of the
ratio of the $TDCS(SRCBA)/TDCS(NRCBA)$ where now the maxima of
this ratio correspond nearly to the local minima of the TDCS when
plotted as a function of the angle $\phi_{B}$.
\begin{figure}[ht]
\centering
\includegraphics[angle=0,width=7cm,height=5.8cm]{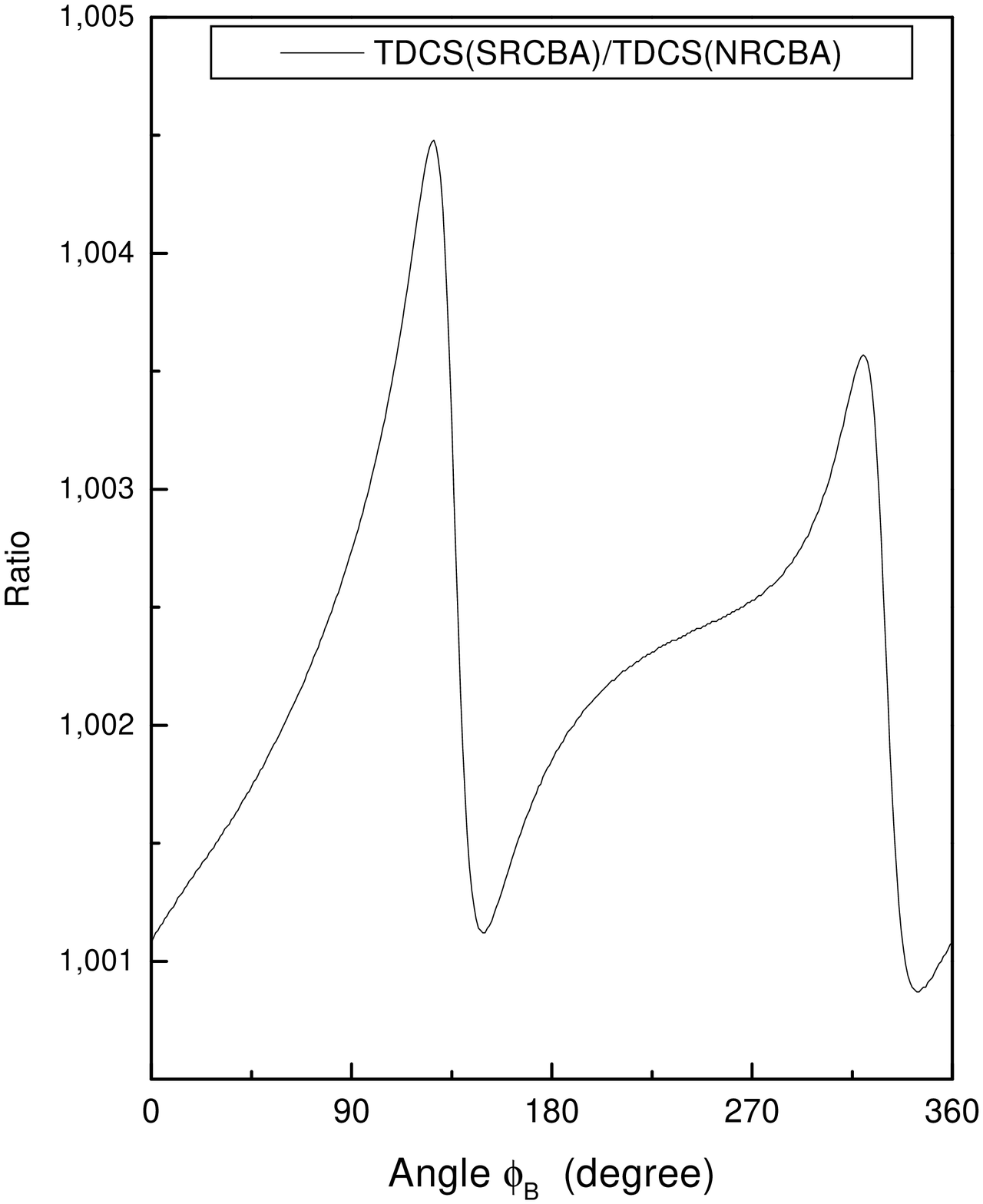}
\caption{The ratio  $TDCS(SRCBA)/TDCS(NRCBA)$ as function of the
angle $\theta_{B}$ with $\theta_{f}=357^{\circ}$. The incident
electron kinetic energy is $T_i=250\quad eV$ and the ejected
electron kinetic energy is $T_B=5\quad eV$}.
\end{figure}
 This is shown in
Fig 4. However, there is no rule that can be inferred from the
behavior of this ratio since when performing various simulations
even in the coplanar asymmetric geometry but with increasing
values of the incident electron kinetic energy, there are many
regions not close to the binary or secondary peaks that present
maxima or minima.
\subsection{Binary coplanar geometries}
The relativistic regime can be defined as follows : when the
value of the relativistic parameter $\gamma =(1-(\beta
/c)^{2})^{-1/2}$ is greater that $1.0053$, there begins to be a
difference between the non relativistic kinetic energy and the
relativistic kinetic energy.
\begin{figure}[h] \centering
\includegraphics[angle=0,width=7cm,height=5.8cm]{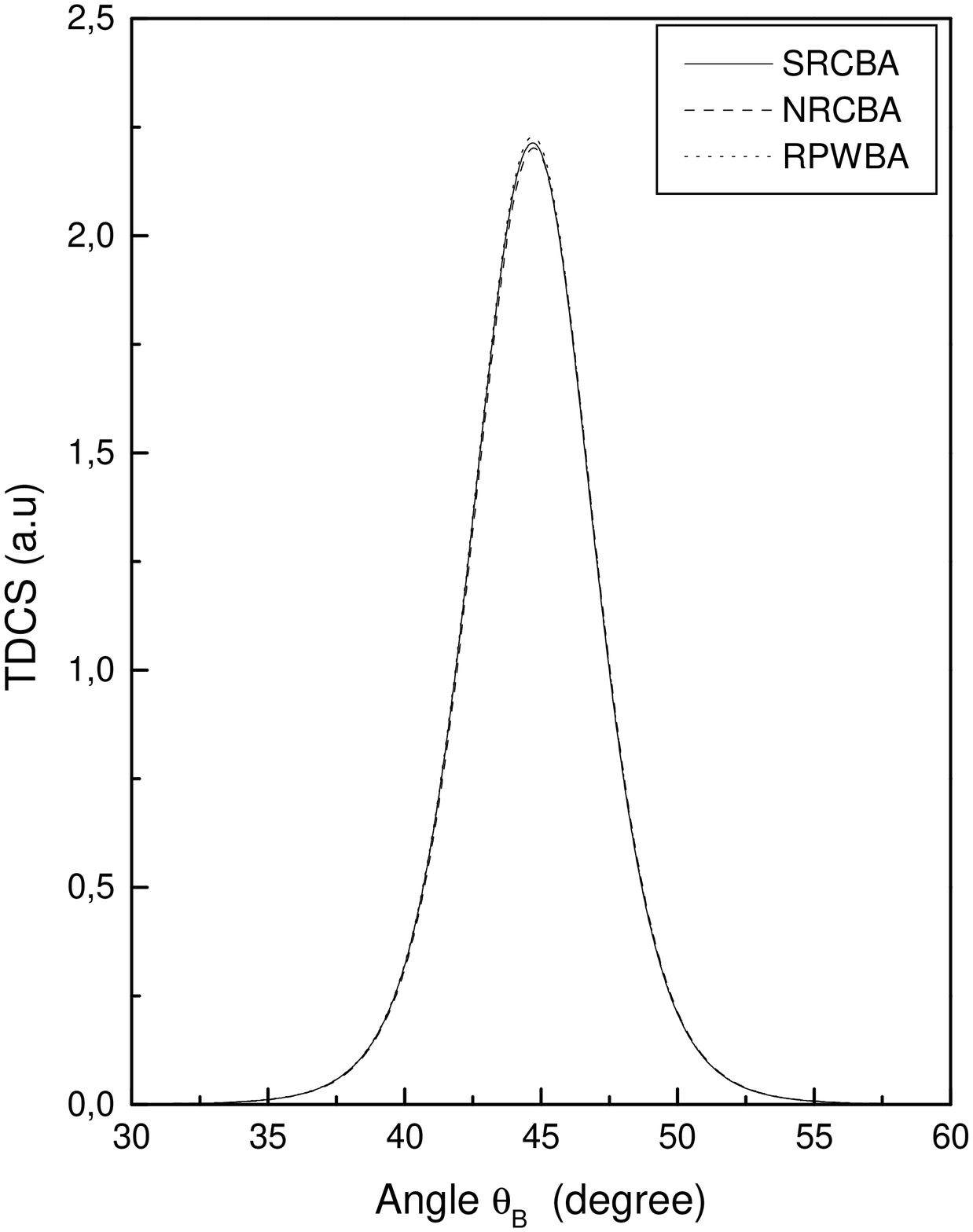}
\caption{The three TDCSs scaled in $10^{-3}$. The solid line
represents the relativistic TDCS in the semi-relativistic Coulomb
Born approximation, the long-dashed line represents the
corresponding TDCS in the non relativistic Coulomb Born
approximation. The short-dashed line represents the rlativistic
plane wave Born approximation. The incident electron kinetic
energy is $T_i=2700\quad eV$ and the ejected electron kinetic
energy is $T_B=1349.5\quad eV$ and $\theta_{f}=45^{\circ}$}.
\end{figure}
This numerical value of the
aforementioned relativistic parameter corresponds to an incident
electron kinetic energy of $E_i=2700\text{ }eV$. Because there is
no experimental data available for this regime, we simply compare
our results with those we have previously found when we
introduced the RPWBA \cite{23} (relativistic plane wave Born
approximation)to study the ionization of atomic hydrogen by
electron impact in the binary geometry. In Fig. 5, it is clearly
visible that the three models (NRCBA, SRCBA and RPWBA) give the
same results which was to be expected since in this geometry, the
use of a Coulomb wave function is not necessary.

\begin{figure}[h] \centering
\includegraphics[angle=0,width=7cm,height=5.8cm]{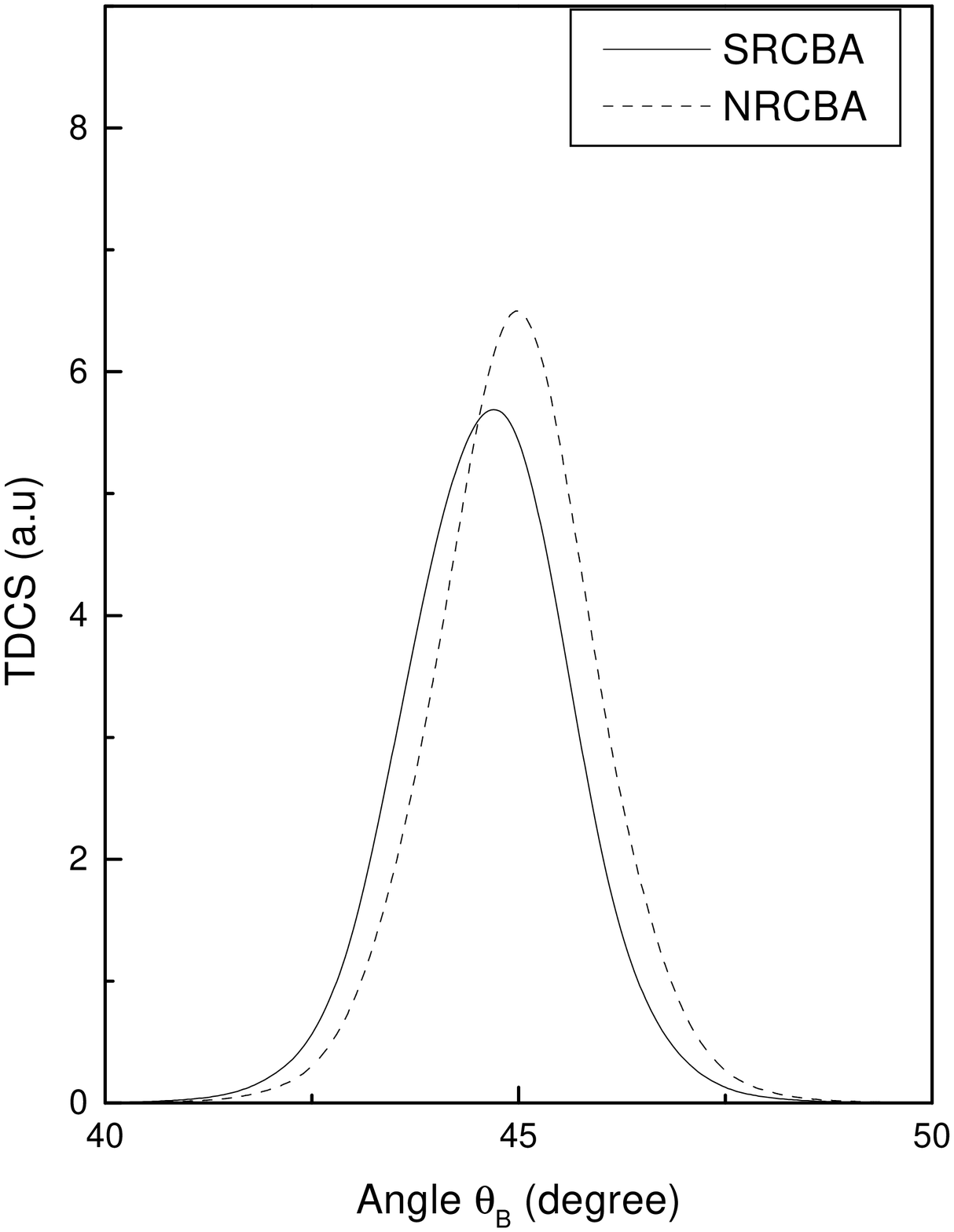}
\caption{The two TDCSs scaled in $10^{-5}$. The solid line
represents the relativistic TDCS in the semi-relativistic Coulomb
Born approximation, the long-dashed line represents the
corresponding TDCS in the non relativistic Coulomb Born
approximation. The incident electron kinetic energy is
$T_i=25000\quad eV$ and the ejected electron kinetic energy is
$T_B=12499.5\quad eV$ and $\theta_{f}=45^{\circ}$}.
\end{figure}
In Fig. 6, there is a shift of the maximum of the TDCS in the
SRCBA towards smaller values than $\theta_B=45^{\circ}$ and this
remains the case for increasing values of the kinetic energy of
the incident electron. The origin of this shift stems from the
fact that the main contribution to the TDCS comes from the term
$H_1(q)$ given by Eq. (16). This term contains a dominant integral
$I_1$. When plotting the behavior of $I_1$ as function of the
angle $\theta_B$, and with increasing values of $E_{i}$, one
observes the shift we have mentioned as well as the fact that in
the relativistic regime, the TDCS(SRCBA) is always lower than the
TDCS(NRCBA).
\section{Conclusion}
In this work, we have developed an exact semi-relativistic model
in the first Born approximation that is valid for a wide range of
geometries, simple in its mathematical structure and that allows
to find previous results using sophisticated non-relativistic
models. This model gives good results if the condition $Z\alpha\ll
1$ is fulfilled.
\appendix
\section{Analytical calculation of the integral $J(\lambda )$}
 Before turning to the analytical
calculation of the integral $J(\lambda )$ proper, let us recall
how the integral $I(\lambda )$ [16] can be obtained. This is
explained without any detail in [20]. Using parabolic
coordinates, one has to evaluate the following integral
\begin{eqnarray}
I(\lambda )&=&\int d\mathbf{r}\exp (i\mathbf{Q.r})\exp (-i\mathbf{p}_{B}.%
\mathbf{r})\frac{e^{-\lambda r}}{r}\nonumber\\
 &\times& _{1}F_{1}(i\eta _{B},1,i(p_{B}r+\mathbf{p}_{B}.\mathbf{r}))
 \end{eqnarray}
The choice of the scalar product $\mathbf{Q.r}$ chosen is [6]
\begin{equation}
\mathbf{Q.r=}\frac{1}{2}Q(\xi -\eta )\cos \gamma -4\sqrt{\xi \eta
}\cos \varphi \sin \gamma
\end{equation}
Performing the various integrals, one finds
\begin{eqnarray}
I(\lambda )&=&\frac{2\pi }{\lambda -i(Q\cos \gamma
+p_{B})}\int_{0}^{\infty }d\xi \exp (-\mu \xi )\nonumber\\
&\times& _{1}F_{1}(i\eta _{B},1,ip_{B}\xi )
\end{eqnarray}
We use the well known result [20]
\begin{equation}
\int_{0}^{\infty }dt\exp (-\lambda t)_{1}F_{1}(\alpha
,1,kt)=\lambda ^{\alpha -1}(\lambda -k)^{-\alpha }
\end{equation}
with
\begin{equation}
\lambda =\mu =\frac{Q^{2}\sin ^{2}\gamma }{2[\lambda -i(Q\cos \gamma +p_{B})]%
}+\frac{1}{2}[\lambda +i(Q\cos \gamma +p_{B})]
\end{equation}
 and $\alpha =i\eta _{B}$ and $k=ip_{B}$. This gives
the result :
\begin{eqnarray}
&&I(\lambda )=\frac{4\pi }{(Q^{2}+\lambda ^{2}+p_{B}^{2}+2Qp_{B}\cos \gamma )}%
\nonumber\\
&\times&\exp \left[ i\eta _{B}\ln (\frac{Q^{2}+\lambda
^{2}+p_{B}^{2}+2Qp_{B}\cos \gamma }{Q^{2}+\lambda
^{2}-p_{B}^{2}-2i\lambda p_{B}})\right]
\end{eqnarray}
To recover the integral $I(\lambda )$ given in EQ. (19) of the
text, one has to make the following substitutions :
\begin{equation}
\left.
\begin{array}{c}
-\mathbf{Q}=\mathbf{q}+\mathbf{p}_{B} \\
Qp_{B}\cos \gamma =\mathbf{Q.p}_{B}=-\mathbf{q.p}_{B}-p_{B}^{2}
\end{array}
\right.
\end{equation}
It is then straightforward to find that
\begin{equation}
\left.
\begin{array}{c}
Q^{2}+\lambda ^{2}+p_{B}^{2}+2Qp_{B}\cos \gamma =q^{2}+\lambda ^{2} \\
Q^{2}+\lambda ^{2}-p_{B}^{2}-2i\lambda p_{B}=q^{2}+\lambda ^{2}+2\mathbf{q.p}%
_{B}-2i\lambda p_{B}
\end{array}
\right.
\end{equation}
so that
\begin{equation}
I(\lambda )=\frac{4\pi }{(q^{2}+\lambda ^{2})}\exp \left[ i\eta _{B}\ln (%
\frac{q^{2}+\lambda ^{2}}{q^{2}+\lambda
^{2}+2\mathbf{q.p}_{B}-2i\lambda p_{B}})\right]
\end{equation}
To calculate
\begin{eqnarray}
J(\lambda )&=&\int d\mathbf{r}\exp (i\mathbf{Q.r})\exp (-i\mathbf{p}_{B}.%
\mathbf{r})\frac{e^{-\lambda r}}{r} \nonumber
\\ &\times&\text{ }_{1}F_{1}(i\eta _{B}+1,2,i(p_{B}r+\mathbf{p}_{B}.\mathbf{r}))
\end{eqnarray}
one uses the same procedures to obtain
\begin{eqnarray}
J(\lambda ) &=&\frac{2\pi }{\lambda -i(Q\cos \gamma
+p_{B})}\int_{0}^{\infty }d\xi \exp (-\mu \xi )\nonumber\\
&\times& _{1}F_{1}(i\eta _{B}+1,2,ip_{B}\xi )  \nonumber \\
&=&\frac{2\pi }{\lambda -i(Q\cos \gamma +p_{B})}\frac{1}{\mu
}\nonumber\\ &\times& _{2}F_{_{1}}(i\eta
_{B}+1,1,2,\frac{ip_{B}}{\mu })
\end{eqnarray}
Performing the various substitutions, one gets the following new
(as far as we know) analytical integral
\begin{widetext}
\begin{eqnarray}
J(\lambda )=\int d\mathbf{r}\exp (i\mathbf{q.r})\frac{e^{-\lambda r}}{r}%
\, _{1}F_{1}(i\eta _{B}+1,2,i(p_{B}r+\mathbf{p}_{B}.\mathbf{r}))
=\frac{4\pi }{(q^{2}+\lambda ^{2})}\text{ }_{2}F_{1}\left(i\eta _{B}+1,1,2,-2%
\frac{[\mathbf{q.p}_{B}-i\lambda p_{B}]}{q^{2}+\lambda
^{2}}\right)
\end{eqnarray}
\end{widetext}
 We have tested this analytical result by performing
the integral using two gaussian quadratures because we have
assumed without loss of generality both $\mathbf{q}$ and
$\mathbf{p}_{B}$ to be parrallel to the $Oz$ axis. The first one,
a Laguerre gaussian quadrature (32 points) to integrate over the
radial variable $r$ and the second one, using a Legendre gaussian
quadrature (32 points) to integrate over the angular variable
$\theta $. The agreement between the analytical result and the
numerical result is excellent. To illustrate this point, we give
as an exemple the results obtained by the two
methods for the following random values of the relevant parameters. For $%
\lambda =1$ ,$\left| \mathbf{q}\right| $=1.015055, $\left|
\mathbf{p}\right| =$0.1055098. The exact result is
\begin{equation}
J_{exact}(\lambda )=(0.57355896,0.12458510)
\end{equation}
and the numerical result is :
\begin{equation}
J_{num}(\lambda )=(0.57355899,0.12458507)
\end{equation}


\begin{thebibliography}{99}


\bibitem{1}W. Nakel and C.T.
Whelan, Phys. Rep., \textbf{315}, 409, (1999).

\bibitem{2}C .T. Whelan, R.J.
Allan, J. Rasch, H.R.J. Walters, X. Zhang, J. R\"{o}der, K. Jung,
H. Ehrhardt, Phys. Rev A, \textbf{50}, 4394, (1994).

\bibitem{3}P. Marchalant, C.T.
Whelan, H.R.J.\ Walters, J. Phys. B, \textbf{31}, 1141, (1998).

\bibitem{4}I. Fuss, J. Mitroy,
B.M. Spicer, J. Phys. B, \textbf{15}, 3321, (1982).

\bibitem{5}I.E.\ McCarthy, E.
Weigold, Cont. Phys., \textbf{35}, 377, (1994).

\bibitem{6}F. Bell, J. Phys. B,
\textbf{22}, 287, (1989).

\bibitem{7}A. Cavaldi, L. Avaldi,
Nuovo Cimento Soc. Ital. Fis., D \textbf{16}, 1, (1994).

\bibitem{8}H. Ehrhardt, Comments
At. Mol. Phys., \textbf{13}, 115, (1983).

\bibitem{9}H. Ehrhardt, K. Jung,
G. Kn\"{o}th, P. Schlemmer, Z. Phys., D \textbf{1}, 3, (1986).

\bibitem{10}J.N. Das, A.N. Konar,
J. Phys. B, \textbf{7}, 2417, (1974).

\bibitem{11}J.N. Das, S.
Chakraborty, Phys. Lett. A, \textbf{92}, 127, (1982).

\bibitem{12}B.L. Mo\"{i}seiwitsch,
Prog. At. Mol. Phys., \textbf{16}, 281, (1980).

\bibitem{13}D.H.\ Jakuba\ss
a-Amundsen, Z. Phys., D \textbf{11}, 305, (1989).

\bibitem{14}D.H.\ Jakuba\ss
a-Amundsen, Phys. Rev A, \textbf{53}, 2359, (1996).

\bibitem{15}L.U. Ancarani, S.
Keller, H. Ast, C.T. Whelan, H.R.J. Walters, R.M. Dreizler, J.
Phys. B, \textbf{31}, 609, (1998).

\bibitem{16}H.S.W.\ Massey and
C.B.O Mohr, Proc. Roy. Soc. A \textbf{140}, 613, (1933)

\bibitem{17}F.W. Jr Byron and
C.J.\ Joachain, Phys. Rep. \textbf{179}, 211, (1989).

\bibitem{18}J. Eichler and W.E.
Meyerhof, \textit{Relativistic Atomic Collisions}, Academic
Press, (1995).

\bibitem{19}W. Greiner and J.
Reinhardt, \textit{Quantum Electrodynamics}, Springer-Verlag,
(1992).

\bibitem{20}L. Landau and E.
Lifchitz, \textit{M\'ecanique quantique}, Editions Mir, Moscou,
(1967).

\bibitem{21}J. Berakdar, Phys. Rev A, \textbf{56}, 370, (1997).

\bibitem{22}J.S. Briggs, Comments At. Mol. Phys., \textbf{23}, 155, (1989).

\bibitem{23}Y. Attaourti and S. Taj, Phys. Rev. A \textbf{69}, 063411
(2004).
\bibitem{24}H. Ehrhardt, G. Knoth, P. Schlemmer and K. Jung, Phys.
Lett. A \textbf{110}, 92, (1985).

\end{thebibliography}
\end{document}